\documentclass[draftcls,onecolumn]{IEEEtran}

\usepackage{graphicx}

\begin{document}
\title {Simulation of hydrogenated graphene Field-Effect Transistors through a multiscale approach}

\author{\normalsize Gianluca Fiori, S. Leb\`egue$^*$, A. Betti, P. Michetti$^+$, M.~Klintenberg$^\dag$, O.~Eriksson$^\dag$, Giuseppe Iannaccone\\
 Dipartimento di Ingegneria dell'Informazione~: Elettronica, Informatica, Telecomunicazioni,\\
Universit\`a di Pisa, Via Caruso, 56126 Pisa, Italy.\\
$^*$Laboratoire de Cristallographie, R\'esonance Magn\'etique et Mod\'elisations (CRM2, UMR CNRS 7036) \\
Institut Jean Barriol, Nancy Universit\'e BP 239, Boulevard des Aiguillettes 54506 Vandoeuvre-l\`es-Nancy, France\\
$^+$Institute for Theoretical Physics and Astrophysics,\\
University of Wuerzburg, D-97074 Wuerzburg, Germany\\
$^\dag$Department of Physics and Astronomy, \\
Uppsala University, Box 516, SE-75120, Uppsala, Sweden\\
email : gfiori@mercurio.iet.unipi.it; Tel. +39 050 2217596; Fax : + 39 050 2217522\\}
\maketitle
\bibliographystyle{IEEEtran}
\begin{abstract}

In this work, we present a performance analysis of Field Effect Transistors
based on recently fabricated 100\% hydrogenated graphene (the so-called graphane) 
and theoretically predicted semi-hydrogenated graphene (i.e. graphone).
The approach is based on accurate calculations of the energy bands 
by means of GW approximation, 
subsequently fitted with a three-nearest neighbor (3NN) sp$^3$ tight-binding Hamiltonian, and finally
used to compute ballistic transport in transistors based on functionalized graphene.

Due to the large energy gap, the proposed devices have many of the advantages 
provided by one-dimensional graphene nanoribbon FETs, such as large $I_{\rm on}$ and
 $I_{\rm on}/I_{\rm off}$ ratios, reduced band-to-band tunneling, 
without the corresponding disadvantages in terms of prohibitive 
lithography and patterning requirements for circuit integration. 

\end{abstract}


Chemical functionalization is a viable route towards
band gap engineering of graphene-based materials, as first demonstrated in~\cite{Elias09},
where exposure to a stream of hydrogen atoms has led to 100\% hydrogenation
of a graphene sheet, the so-called {\it graphane}.

Obviously, research on functionalized graphene devices is at an embryonic stage.
Several issues must be addressed to introduce graphane in future generations of electron devices. 
From this perspective, theoretical simulations can be very useful to explore possible solutions for
device fabrication and design, in order to provide an early assessment of the opportunities of
functionalized graphene in nanoscale electronics.

Sofo et al.~\cite{Sofo07} first predicted stability of 100\% hydrogenated graphene 
 through standard DFT calculations with a Generalized Gradient Approximation (GGA). 
As also observed in~\cite{Boukhvalov08}, the H atom adsorption leads to sp$^3$ hybridization, 
where three of the sp$^3$ bonds are saturated by C atoms, and the fourth by H atom, which in turn
induces an energy gap opening of few eVs. More detailed simulations have been performed in~\cite{Lebegue09}, 
where DFT deficiencies in calculating energy gap have been overcome through GW simulations, showing energy gaps 
of 5.4 eV and 4.9 eV for the chair and the boat conformation, respectively, 
thus correcting results in~\cite{Sofo07, Boukhvalov08} and~\cite{Nakamura09} by almost 2 eV. 
Non-ideal structures have been studied in~\cite{Flores09}, through geometry optimization 
and molecular dynamics simulations. 
It has been shown that H frustration is very likely to occur, 
leading to extensive membrane corrugation, but also that hydrogenated domains,
once formed, are very stable.
Atomistic simulations have demonstrated that semi-hydrogenated graphene ({\it graphone}) with
$H$ atoms on the same side, possesses ferromagnetic properties, opening graphene to spintronic applications~\cite{Zhou09}. 

Ferroelectric behavior has also been determined in~\cite{Singh09}, where ``nanoroads'' 
(i.e. graphene nanoribbons) have been defined on fully hydrogenated carbon sheets, 
exhibiting the same energy gap behavior as a function of width as in~\cite{Son06}.

Gap opening can be induced not only through H-functionalization, but also by means of other absorbants~\cite{Leb10}. 
Fluorine has been taken into account in~\cite{Lu09}, demonstrating a clear dependence of the energy gap on fluorine 
concentration, as well as lithium~\cite{Yang09}, which however presents a geometrical conformation different 
from that in graphane, in which C atoms are pulled out by absorbants.

All mentioned articles are concerned with simulations of material properties, whereas studies 
on the operation and performance of functionalized graphene-based devices are lacking, except for one work 
on current-voltage characteristics of graphane p-n junctions~\cite{Gharekhanlou09}, based 
on the effective mass approximation.
Clearly, more efforts have to be directed towards this direction, since whether graphane is suitable
as a channel for field effect transistors is still an open issue, which can benefit from 
a contribution based on accurate numerical simulations.\cite{IannaIEDM2009}

In order to address all these issues and to advance research in graphane electronics, 
we present a multiscale approach, based on {\sl i)} accurate GW calculations of the energy dispersion relations,
{\sl ii)} a fitting of the computed energy bands by means of a 3NN sp$^3$ tight-binding Hamiltonian to be included in a {\sl iii)}
a semiclassical model, based on the assumption of ballistic transport, to simulate Field-Effect Transistors based on graphane and graphone channels.

\begin{figure} [tbp]
\vspace{0.4cm}
\begin{center}
\includegraphics[width=13.1cm]{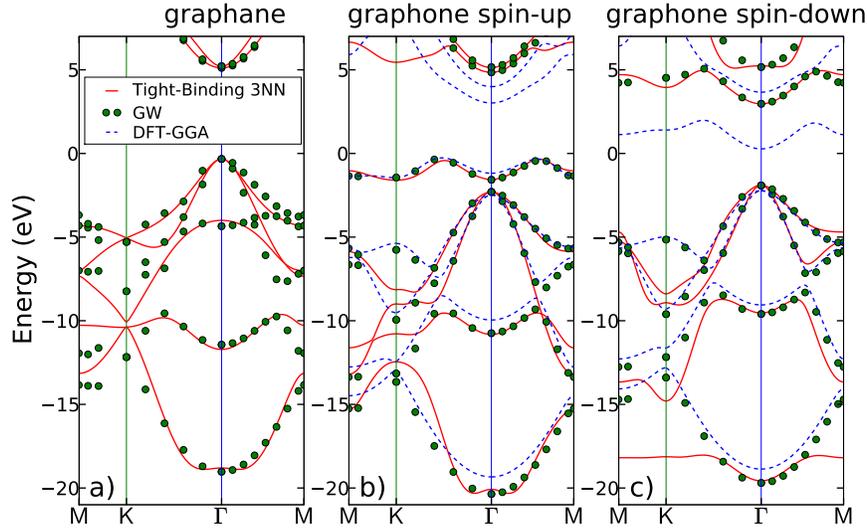}
\end{center}
\caption{Energy bands computed for graphane (a) and graphone (b-c), using tight-binding three-nearest neighbor (solid line),
GW (dots) and DFT-GGA approaches (dashed lines). In the case of graphone spin-up (b) and
spin-down bands (c) are shown. }
\label{Fig1}
\end{figure}

In Fig.~\ref{Fig1}, we show the computed energy bands by means of three different models: three-nearest neighbor
sp$^3$ tight-binding Hamiltonian (solid line), DFT within generalized gradient approximation (GGA) 
of Perdew Burke Ernzerhof\cite{PBE} (dashed line) and GW (dots)\cite{Hedin1,Hedin2}.
DFT calculations have been performed by means of the Vienna Ab-initio Simulation Package\cite{Kresse1,Kresse2}, which implements
density functional theory\cite{Hohenberg,Kohn} in the framework of the projector augmented waves method\cite{Blochl}. 
Due to the presence of an unpaired electron in graphone\cite{Zhou09}, we have allowed for spin-polarization in all our calculations.
In addition, since a ferromagnetic arrangement of the magnetic moments gives the smallest total energy\cite{Zhou09},
we have considered the simplest cell that can accommodate this magnetic order, composed by two carbon and one
hydrogen atoms.

Since the GGA approach is not suitable to treat excited states and the band gap, which is usually underestimated by
standard exchange-correlation functionals such as the LDA or GGA, we have adopted the GW approximation (GWA).

In particular, we have exploited the GW approximation as implemented in the code VASP\cite{gwkresse}, 
which provides similar results as those from an earlier implementation\cite{Leb03}.
In order to help convergence, we have considered 200 bands for the corresponding summation in the calculation 
of the polarizability and the self-energy, while we have used a cut-off of $150$ eV for the size of the polarizability matrices.

Since a comparison between DFT-GGA and GW in graphane has been extensively investigated in a previous 
article by some authors of the present paper\cite{Lebegue09}, here we just summarize the main features of its electronic structure. 
In the chair conformation, which is the most stable\cite{Sofo07}, 
the gap is direct and in correspondence of the $\Gamma$ point, and the bandgap is equal to 3.5 eV for the GGA, 
and 5.4 eV\cite{Lebegue09} for the more precise GW approximation. In the same way, the transitions at the high-symmetry
points $M$ and $K$ are increased when considering the GW: from 10.8 eV to 13.7 eV at the $M$ point, and
from 12.2 to 15.9 eV at the $K$ point\cite{Lebegue09}. The GW bandstructure of graphane presented here is exactly
equivalent to the one in~\cite{Lebegue09}, and it is used here as a support for the tight-binding fitting.

In Fig.~\ref{Fig1}b and Fig.~\ref{Fig1}c, we show spin-up and spin-down energy bands computed by means of
 the three above mentioned models for the 50\% functionalized graphene sheet (graphone).
The valence band maximum is located along the $\Gamma-K$ high-symmetry line (Fig.~\ref{Fig1}b), and
almost degenerate with the other maximum along the $\Gamma-M$ direction. 
The conduction band minimum is instead at the $\Gamma$ point (Fig.~\ref{Fig1}c).
For both GW and GGA the gap is indirect. However, while GGA provide a bandgap equal to 0.46~eV, the GWA 
provides a bandgap equal to 3.2 eV. 
In addition to increasing the band gap, the GWA is also modifying the dispersion relation with respect to GGA.

In order to solve the electrostatic and the transport problems in functionalized graphene devices, 
we have adopted a semiclassical model similar to that in~\cite{Kos}, able to compute the free charge density 
and the current, given the energy bands in the whole Brillouin Zone (BZ).
The choice for such a model is justified by the large energy gap obtained for both graphane and graphone, 
which strongly limits the band-to-band tunneling component, differently to what happens in carbon nanotubes~\cite{NoiCNT}, 
graphane nanoribbon~\cite{NoiGNR} or graphene bilayer transistors~\cite{Noibilayer}.
The model relies on the assumption of fully ballistic transport, since our aim in the current
work is to assess the upper limits of device performance.

From a numerical point of view, a large number of $k$ points in the BZ (almost equal to 10$^5$)
is required in order to obtain accurate results, which can be really prohibitive for the GW approach. 
In order to avoid this problem, the energy dispersion along the $\Gamma$M and $\Gamma$K 
directions obtained by the GW approximation, has been fitted by means of a least-mean square procedure 
and a three-nearest neighbor sp$^3$ tight-binding Hamiltonian,
 which has demonstrated to provide a better fitting as compared to a simple nearest-neighbor approach.
Results are shown in Fig.~\ref{Fig1}. Particular attention has been paid to
 the minimum and the maximum of the conduction and valence bands,
respectively, since such states are those mainly contributing to transport.
As can be seen, tight-binding results are in good agreement with GW calculations. 

\begin{figure} [tbp]
\vspace{0.4cm}
\begin{center}
\includegraphics[width=9cm]{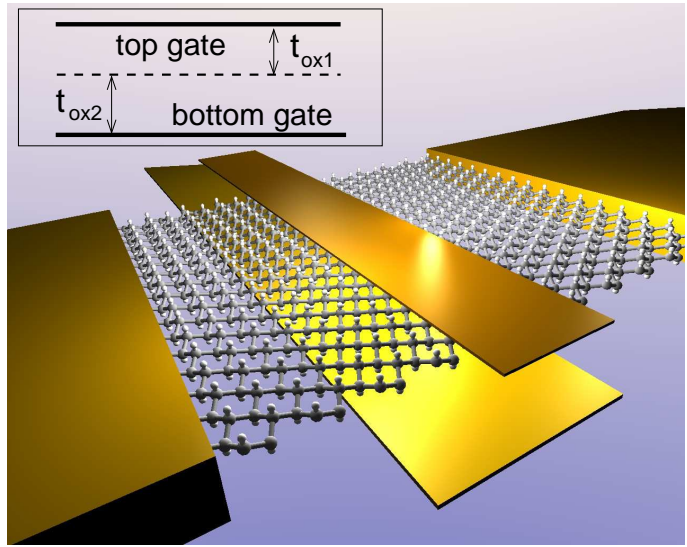}
\end{center}
\caption{Sketch of the simulated device. The channel here shown is graphane, but the very
same structure has been considered for graphone based FET. In the inset, the device transversal cross-section is shown.}
\label{Fig2}
\end{figure}

In Fig.~\ref{Fig2}, we show the structure of the simulated double gate device. The channel is embedded in SiO$_2$ ($\varepsilon_r$=3.9).
In the inset of Fig.~\ref{Fig2}, we show the device transversal cross-section: 
$t_{\rm ox1}$ and  $t_{\rm ox2}$ are the top and bottom gate oxide thicknesses, respectively.

\begin{figure} [tbp]
\vspace{0.4cm}
\begin{center}
\includegraphics[width=15cm]{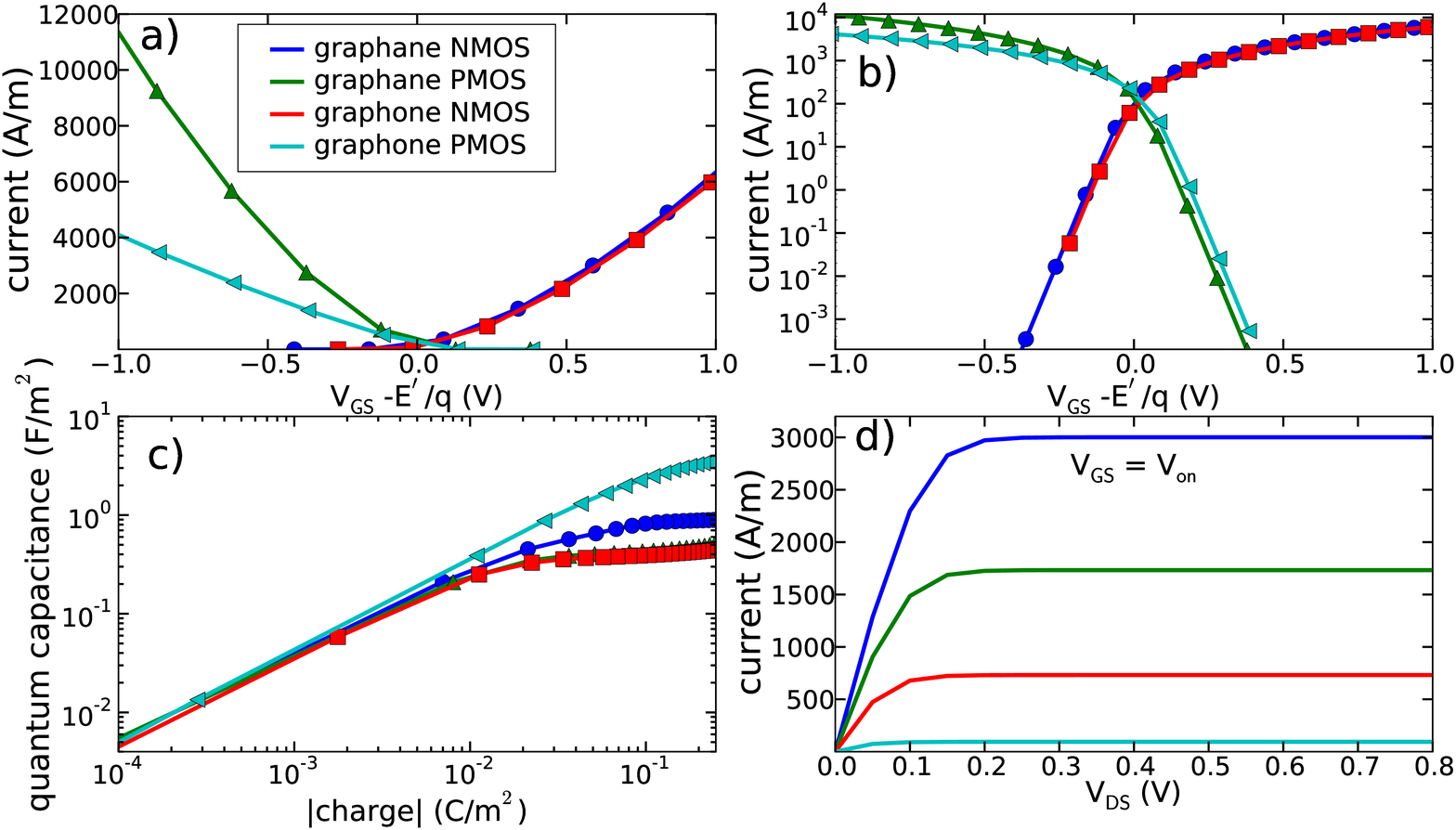}
\end{center}
\caption{Transfer characteristics for graphane and graphone NMOS and PMOS in the linear a) and in the logarithmic scale b): 
each transfer characteristic has been translated by $E^\prime$, the bottom of the conduction band for NMOS and 
the top of the valence band for PMOS;
 c) quantum capacitance as a function of charge density in the channel; 
d) output characteristic for different $V_{\rm GS}$: the gate-to-source step is 0.2~V.}
\label{Fig3}
\end{figure}

In Fig.~\ref{Fig3}a and Fig.~\ref{Fig3}b, the transfer characteristics of NMOS and PMOS FETs based on graphane and graphone are shown in the linear
and logarithmic scale, for $t_{\rm ox1}$=$t_{\rm ox2}$=1~nm, and for a drain-to-source voltage $V_{\rm DS}$=$V_{\rm DD}$=0.8~V.
 To allow a comparison with the scaling expected for silicon technology, 
we consider the value of $V_{\rm DD}$ that the International Technology Roadmap for Semiconductors 
~\cite{ITRS} predicts for high performance logic in 2015-2016. Analogously, the current in the OFF state $I_{\rm off}$ has been set to 100~nA/$\mu$m.

In the sub-threshold regime, as expected, the sub-threshold swing (SS) is equal to 
60~mV/dec, due to the adopted double gate geometry, which assures good gate control over
the channel barrier.

As can be seen, all devices are able to provide large currents and almost 
present similar transfer-characteristics as well as the same transconductance
(derivative of the transfer characteristic with respect to gate-to-source voltage ($V_{\rm GS}$)), except
for graphone PMOS, which shows degraded performance. This can be explained by its larger
quantum capacitance ($C_Q$) (Fig.~\ref{Fig3}c), where $C_Q$  for the four considered devices
 is depicted as a function of the charge density in correspondence of the channel.
As can be seen, in the above threshold regime (charge density larger than $10^{-2}$~C/m$^2$), 
the graphone PMOS shows large $C_Q$,
 greater than the electrostatic capacitance ($C_{\rm el}$=6.9$\times 10^{-2}$~F/m$^2$).

From Fig.~\ref{Fig3}c, one can also extract information concerning the effective mass.
Quantum capacitance is indeed proportional in the flat region to the effective mass of the particle in the
2DEG~\cite{Luryi}. As also confirmed by the curvature of the bands (which is inversely proportional to the effective mass),
lighter particles are those in correspondence of the valence band of graphane and of the conduction band of graphone,
while particles in graphone valence band are the heaviest ones.

In Fig.~\ref{Fig3}d we show the drain-to-source current as a function of $V_{\rm DS}$: the $V_{\rm GS}$ step is equal to 
0.2~V.

\begin{figure} [tbp]
\vspace{0.4cm}
\begin{center}
\includegraphics[width=15cm]{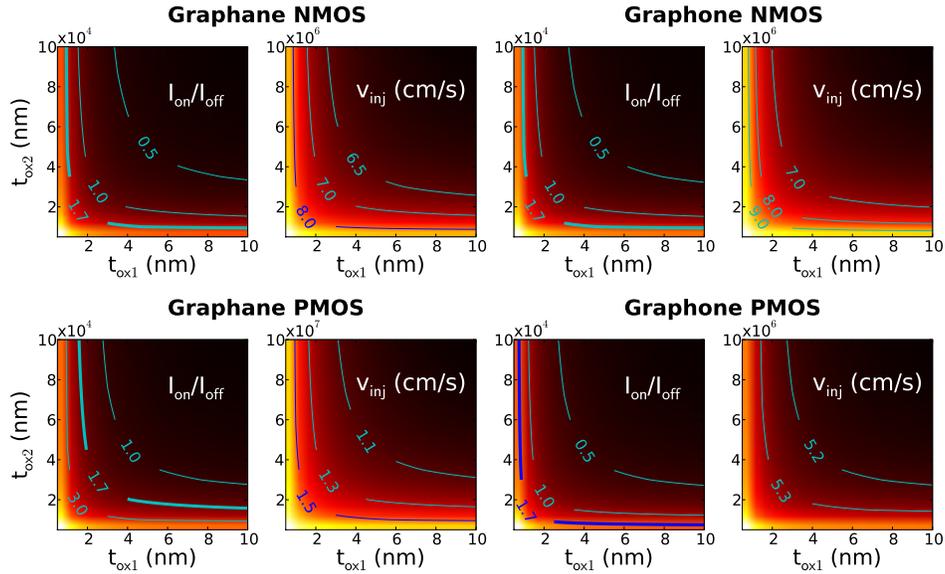}
\end{center}
\caption{Colormap of $I_{\rm on}/I_{\rm off}$ ratios and injection velocity for graphane and graphone NMOS and PMOS devices
as a function of $t_{\rm ox1}$ and $t_{\rm ox2}$.}
\label{Fig4}
\end{figure}

Fig.~\ref{Fig4} shows the $I_{\rm on}/I_{\rm off}$ ratio and the injection velocity for the four considered devices
as a function of the top and bottom oxide thickness. Since $I_{\rm off}$ is fixed ($I_{\rm off}$=100~nA/$\mu$m),
$I_{\rm on}$ can be directly extracted from the colormaps.
The isolines for $I_{\rm on}/I_{\rm off}$=1.7$\times 10 ^4$, 
which is the ratio required by the ITRS for the 2015-2016 technology, are also highlighted. As can be seen, 
all the devices manage to provide 
large ratios, even when considering top gate oxide thicknesses of almost 1~nm, and large bottom gate oxide thickness.

Injection velocity ($v_{\rm inj}$) has been instead computed in correspondence of $V_{GS}=V_{\rm on}$, 
defined as $V_{\rm on}=V_{\rm off}+V_{\rm DD}$,
where $V_{\rm off}$ is the $V_{\rm GS}$ at which the current is equal to $I_{\rm off}$ and represents the velocity of  
thermally emitted electrons from the reservoirs to the channel.

Holes in graphone PMOS show the slowest velocity, in accordance with the above considerations, 
while the fastest particles are holes in graphane  PMOS. 

In conclusion, a multi-scale approach has been adopted in order to assess potential of functionalized graphene
as channel material for next-generation high-performance Field-Effect Transistors.
To achieve this task, calculations within the GW approximation have been performed
in order to compute accurate energy bands and band-gaps, which, in the case of graphone, has been demonstrated to 
differ by more than 2.7~eV from previous results.
Tight-binding Hamiltonians for graphane and graphone have been fitted with DFT results
in order to feed a semiclassical model, able to compute transport in the whole Brillouin Zone.
Results have shown that graphane and graphone based FETs can provide large current as well as 
$I_{\rm on}/I_{\rm off}$ ratios, and can represent a promising option for future technology nodes.

{\bf Acknowledgment} - The work was supported in part by the EC Seventh Framework
Program  under the Network of Excellence NANOSIL (Contract 216171) and the STREP 
project GRAND (Contract 215752). S. L. acknowledges financial support from ANR PNANO 
grant N$^0$ ANR-06-NANO-053-02 and computer time using HPC resources from GENCI-CCRT/CINES 
(Grant 2010-085106). M.K. acknowledges financial support from the Swedish Research Council (VR), 
and the G\"oran Gustafsson Stiftelse. O.E. acknowledges VR and ERC for support.

\bibliography{article}

\end{document}